# Study of the exfoliation and functionalization of graphene from graphite flakes with plasma discharge in solution

LIEBGOTT Q., DA SILVA TOUSCH C., LETOFFE A., IBRAHIM D., KABBARA H., NOEL C., HENRION G., HEROLD C., ROYAUD I., PONCOT M., FONTANA S., CUYNET S.

IJL, Université de Lorraine, CNRS, 2 allée André Guinier, F-54000 Nancy, France

**Abstract:** Graphene flakes were produced by nanosecond plasma discharge at atmospheric pressure between an electrode and the surface of distilled water, in which were placed graphite flakes. The discharge ionizes the gas and forms free radicals on the surface of the water, functionalizing the graphite flakes in solution. The plasma also gives enough energy to break the Van der Waals bonds between the graphene layers but not enough to break the covalent C-C bonds within the layers. Transmission electron microscopy confirmed the hexagonal structure of graphene sheets, and showed that they were monocrystalline. No contamination was found in the obtained nanomaterial. An unknown phenomenon has been found in the activated distilled water, making its electrical conductivity decrease with an increasing temperature. An acidification of the water is observed. The gas in which the discharge takes place plays a major role on the process, no exfoliation is observed if plasmogen argon gas is used.

## I. INTRODUCTION

Graphene is an allotropic form of carbon, and its structure is a single planar sheet of carbon atoms arranged in a hexagonal pattern. Graphite is the stacking of graphene sheets, bonded together by Van der Waals forces. Over the last 15 years, since its discovery, graphene has attracted much interest from researchers, because of its properties such as an extremely high mechanical strength[1], a high specific surface area[2], a high transparency[3, 4], and a high thermal conductivity[5]. Furthermore, graphene could be an alternative material for electrodes fuel cells, sensors, transparent conductive films, flexible electronics to name some[3, 6, 7, 8].

Since its discovery by *Novoselov et al.*[9] in 2004 with the peel-off method using duct tape, other methods have been developed for its synthesis, such as chemical vapour deposition (CVD) and chemical exfoliation. The major drawback of CVD-growth graphene is the potential contamination by organic compounds[10, 11, 12, 13]. With chemical exfoliation, the graphene sheets obtained are larger but have many defects which lower their conductivity and mechanical properties, limiting their use for large area requiring applications [14, 15, 16]. Another method use for exfoliation of graphene from graphite is ultrasonication in liquid phase[17]. Often, a surfactant is added in the organic solvent, meaning that the obtained nanomaterial needs to be purified after the process. Producing large and defect-free graphene sheets in large quantities currently is a challenge, and graphene's properties decreases significantly with the number of graphene layers stacked and in presence of defects.

Plasma-assisted methods to obtain graphene offer the possibility to prepare graphene at low temperature, possibly decreasing the cost of the processes. Some of these methods do not use solvents nor surfactants, and avoid some contamination of the prepared materials. However, the use of plasma is challenging because the energy delivered needs to be adjusted to break only the Van der Waals bonds and avoid damage caused by the ion bombardment[18,19]. Moreover, the Van der Waals forces are smaller than covalent C-C bonds in the planes. This highlights the importance of the energy given by the plasma, that must exceed the Van der Walls forces without overcoming the covalent C-C bonds. Studies reported that too high energy plasma creates more carbon onions than graphene sheets[20].

In this paper we present a new plasma process generated between an electrode and the surface of distilled water for exfoliation of graphene, which is not time consuming and which allows to prepare contamination-free graphene sheets.

Plasma-liquid interactions are a growing interdisciplinary area of research combining plasma science, plasma-induced chemistry, fluid dynamics, and heat and mass transfer. Thus, recent researches are a basis to analyse the plasma-activated water and knowing the species that could be formed due to the presence of the plasma. The plasma is a partial ionization of a gas phase, and the interaction of it with a liquid can solubilize gaseous species and form free radicals of species found in the liquid. Thus, the final solution contains different species and ions that were not in the original distilled water.

A cold plasma at atmospheric pressure that induces plasma-liquid interactions, forming free radicals in the water, allowing the simultaneous exfoliation and functionalization of graphite flakes.

## II. MATERIALS AND METHODS

A schematic representation of the experimental setup is shown in Fig. 1. Madagascar graphite flakes with 20-100 µm lateral dimensions were used as carbon precursor. One electrode made of graphite is immersed in distilled water, the other electrode made of tungsten is placed 1 mm above the surface of the liquid. The discharges are generated by a generator(Spellman PTV10P350/230)coupled with a switcher(Nanogen 1 Smart HV Pulses Generator), operating at a voltage of 6 kV, a frequency of 6 kHz, a pulse width of 1 µs, between the upper electrode and the surface of the 80 mL distilled water which exhibits a conductivity of 2 µS.cm$^{-1}$ conductivity. In this process, the surface of the liquid acts as an electrode, and the air breakdown voltage will depend on the gap between the upper electrode and the surface on the liquid, which is determined by Paschen's law to be 1.3 mm at atmospheric pressure for a 6 kV voltage. For the case of argon, this gap distance is approximately 8 mm.

Throughout the elaboration, current and voltage were monitored with a current probe (MagneLab CT-D1.0-BNC) and a high-voltage probe (PMK-14KVC PHV 4-3925), respectively. The probes are connected to an oscilloscope (LeCroyWaveSurfer 104 MXs-B). The values given are the values of the voltage and current during the 1 µs pulse, at 900 ns.

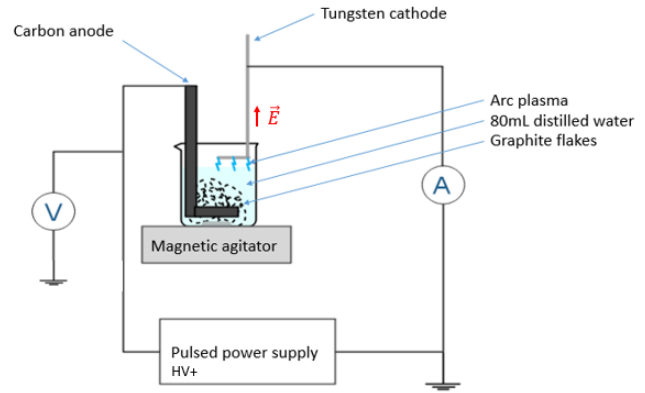

Fig. 1: Representation of the experimental setup.

Because of the natural aggregation of the carbon materials with time after the process, ultrasonication was used to separate them before carrying out the characterizations.

The obtained carbon nanomaterials were analysed used different characterization techniques aiming to analyse different properties of the materials. First, the surface morphology was analysed using SEM (Quanta FEG250) with a 20 kV accelerating voltage. The samples have been prepared by placing a drop of the suspension containing carbon nanomaterials on a silicon water, then drying it by heating. The morphology and crystallinity was further characterized using TEM (Philips CM200) with a 200 kV accelerating voltage. Sample preparation consists in placing a drop of the same suspension on a Cu TEM grid. The microstructure and defects were measured using Raman spectroscopy (HR800 Horiba) with a 633 nm laser at room temperature. Functionalization was measured using thermogravimetric analysis (TGA) (SETARAM Setsys Ev 1750) from 20 °C to 1100 °C, with a slope of 3 °C.min$^{-1}$, coupled with a mass spectrometer (Pfeiffer Vacuum Omnistar GSD 301C). Prior to TGA, the sample has been dehydrated and put in an alumina crucible.

After the process, a diagnosis of the activated distilled water was made using a pH meter (DeltaOhm HD 2165.2), a conductivity and temperature probe (DeltaOhm SP 06 T) and a pH probe (DeltaOhm KP30).

## III. RESULTS

Many diagnoses have been made, for example an electrical diagnosis has been made during the process. After the process, the solution is diagnosed by its pH and its electrical conductivity. The obtained material is characterized, and the plasma is observed.

III.1 Dispersion of the graphite flakes in solution

Before generating the plasma at the surface of the liquid, the graphite flakes are hydrophobic, making them stay on the surface of the liquid, even if the agitation makes them sink, they tend to go back on the surface. After a flake is hit by the plasma discharge, it instantly becomes hydrophilic, and when the agitation stops, large flakes sink, whereas small particles are suspended in solution. The time needed for a graphitic particle to sink depends on its dimensions, meaning the more time it takes to sink, the higher the specific surface area of the particle.

The suspended particles stay in suspension for different times; the first ones aggregate on the bottom of the container in 5 hours after sonication to separate particles, but some particles stay in suspension in the liquid for up to 5 days.

Plasma discharge has an influence on the hydrophilic behaviour of the graphite flakes, since putting graphite flakes in plasma-treated water does not make them hydrophilic, they stay above the surface of the liquid. This highlights the importance of the plasma in this change of behaviour of the particles.

### III.2 Plasma characterization

#### III.2.a Impact of the gas

The plasma discharge is different depending on the gas in which it is generated. In air, after some minutes, the behaviour of the plasma changes. By direct observation it goes from a columnar blue/pink plasma to a filament pink/red plasma. This displays the change of atmosphere in the region of the plasma. As it takes place just above the surface of the liquid, changing from air to air saturated with water vapour.

If pure $N_2$ is used as plasmogen gas, the same observations are made, but delayed in time as the current kinetics are different.

When the discharge is generated in argon, the plasma is columnar and purple. As the current does not increase in argon, the Joule effect is insignificant and the water does not heat up. Thus, the behaviour of the plasma does not change with process time.

Moreover, a plasma is a partial ionization of particles in gas phase. In the case of $N_2$ or air, nitrogen is ionized and when going to the surface of the liquid, can react with species present at the interface between the liquid and the plasma, and solubilise in the liquid. Species formed are mostly ionic. Additionally, given the high energy of the plasma, it is able to dissociate water molecules into free radicals, allowing new species to be formed, such as $H_2O_2$ for example.

However, in argon no species in the liquid during the process will be containing atoms coming from the gas phase.

Further explanation concerning the results obtained with air, $N_2$ and argon will be reported in the following sections of this paper.

#### III.2.b Current kinetics

The current kinetics is important to monitor, as it gives information on the electric parameters throughout the process. Fig. 2 shows the increase of the current during the process. It also shows the decrease of the voltage due to the limited power of the HV generator which is not able to deliver 6 kV at such high current.

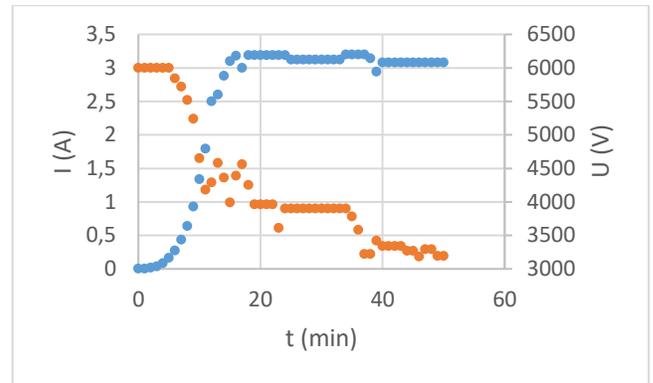

Fig. 2: Evolution of the current (blue) and voltage (orange) with process time in distilled water with air as plasmogen gas.

The maximum current reached during the process in air is 3.19 A, reached at 16 minutes. At this point, the pH has already reached its final value, as well as the electrical conductivity.

The current-voltage kinetic depends on the gap between the electrode and the surface of the liquid, but also on the gas in which the plasma is generated. In Fig. 3 is shown the difference in current kinetics between discharge in air, in $N_2$, and in argon.

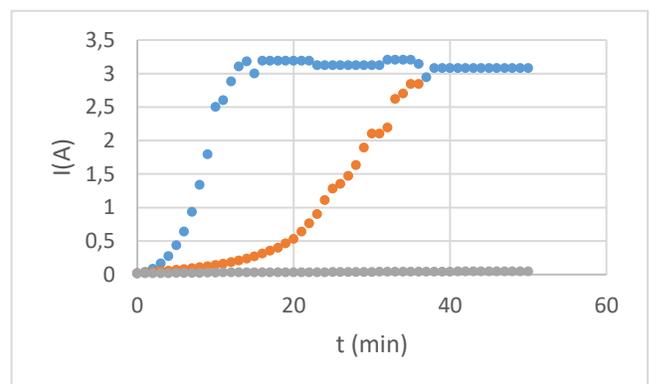

Fig. 3: Current kinetics in function of the gas used for plasma discharge: air (blue), $N_2$ (orange), Ar (grey).

It is interesting to notice that the current did not increase when argon is used for the discharge, no matter the process time. Moreover, using pure N2 gas as plasmogen, an increase in current is observed as for air, but delayed in time.

III.3 Materials characterization

III.3.a Graphite flakes after exfoliation

After the exfoliation process, graphene sheets have been exfoliated from graphite flakes, but as the yield is not high enough to exfoliate the whole flakes into graphene, graphite flakes are still present in the liquid at the end of the process. These flakes have been dried and characterized with SEM, in order to analyse their surface morphology. The surface morphology of the graphite flakes before and after exfoliation are displayed on Fig. 4 and Fig. 5, respectively.

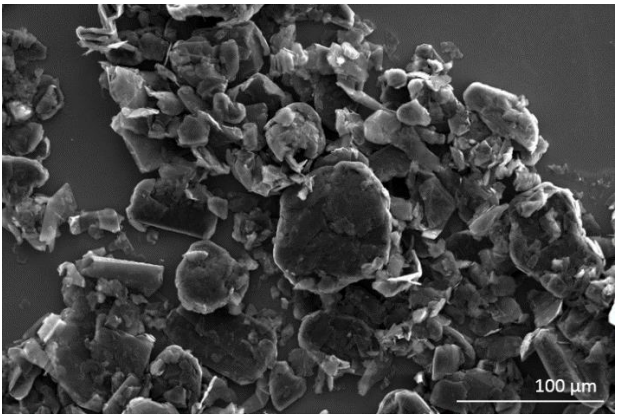

Fig. 4: SEM micrograph of the surface morphology of the graphite flakes before the process.

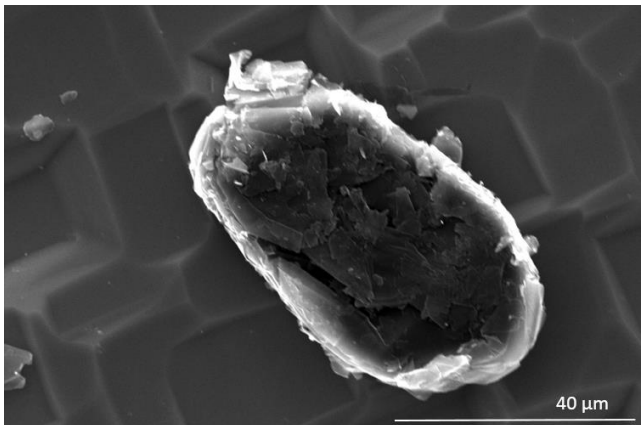

Fig. 5: SEM micrograph of the surface morphology of the graphite flakes after the air process.

What can be seen on Fig. 4 and 5 is also the thickness of the exfoliated sheets on the surface of the flakes. They seem thin enough and exfoliated.

Besides, there seems to be defects on the flakes that were not seen on the untreated material. There are of different types, shown in Fig. 6, such as holes due to the arc roots, fissures under the top layer and on the edge of the particle. However, the cause of this was still unknown, and since the current increases a lot during the process, the main hypothesis was that low and high current discharges (respectively high voltage and low voltage discharges) would have different effects on the materials, thus damaging flakes differently.

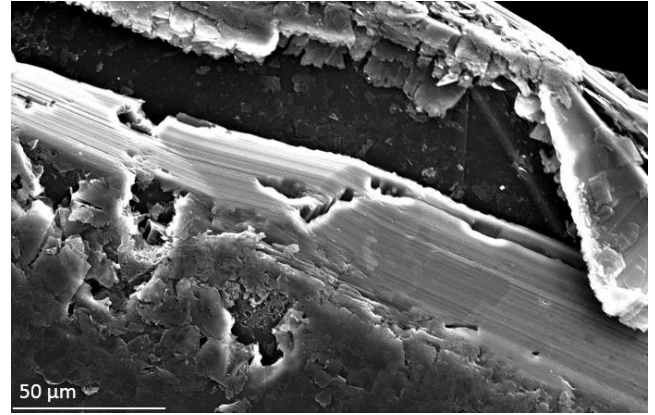

Fig. 6: Defects caused on the graphene flakes during the plasma process.

For this reason, two distinct experiments have been done, to comprehend the action of low and high current during the plasma discharge on the graphite flakes. The first experiment was the same process, but the process length was reduced to 7 minutes, to keep the current between 50 mA and 1 A. The first experiment was done in plasma-treated water, which means the current when the process begins already is 3.8 A, and goes up to 7.2 A in 7 minutes.

The obtained flakes were then characterized using SEM to evaluate their surface condition. The obtained SEM micrographs are presented on Fig. 7 and Fig. 8.

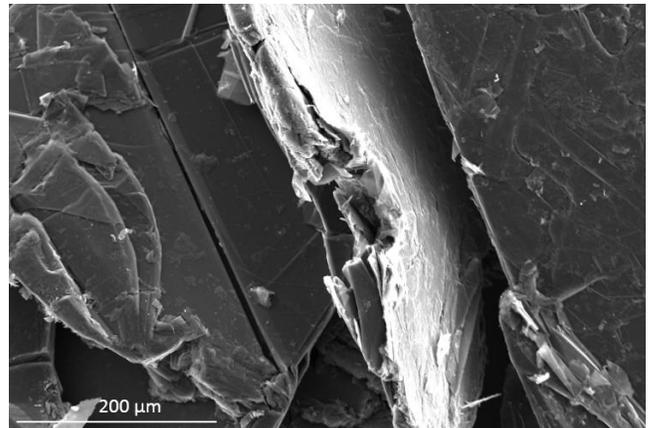

Fig. 7: SEM micrograph of the surface condition of the graphite flakes after the low current experiment.

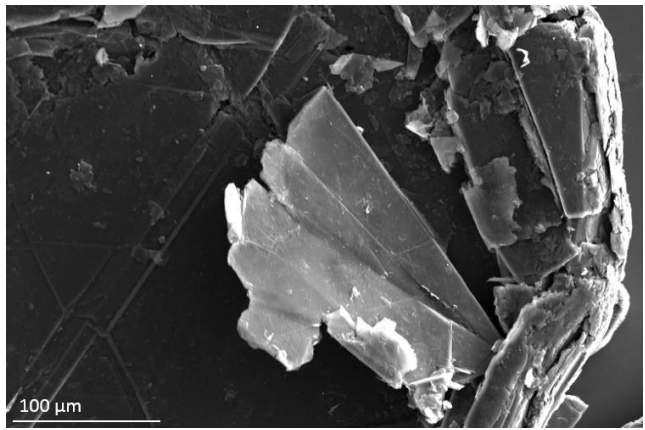

Fig. 8: SEM micrograph of the surface condition of the graphite flakes after the high current experiment.

With high current, holes and fissures in the material due to the process seem more frequent than for low current. Thus, too high current caused by the increase of the water's conductivity could be harmful for the process. However, with the little information we can obtain from the small number of particles observed, further results will be given in the following part to display the effect of the current on the exfoliated particles.

No carbon contamination coming from the immersed carbon electrode was observed with SEM or TEM, and its mass did not reduce during the process.

III.3.b  Graphene sheets

TEM analysis was used to determine the morphology, the crystallinity and the lateral dimensions of graphene. Another important aspect that could have been investigated is the thickness of the graphene sheets.

Nevertheless, qualitative information can still be recorded. As an example, Fig. 9 and Fig. 10 display a representative particle found on the TEM grid, for the process in air and in argon respectively, and their SAED patterns.

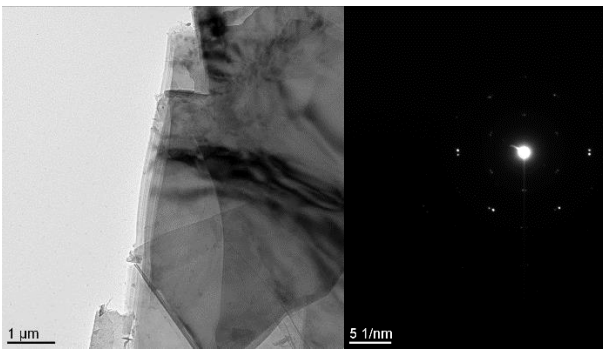

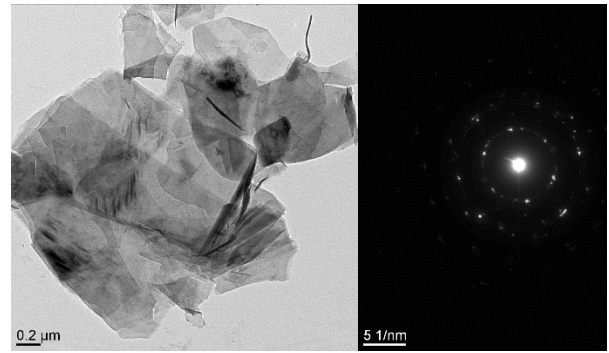

Fig. 9: TEM micrographs of a graphene sheet for the process in air (left) and its SAED (right) pattern.

Fig. 10: TEM micrographs of a graphene sheet for the process in argon (left) and its SAED (right) pattern.

The results for the process in air are encouraging since large monocrystalline particles have been obtained. However, for the process in argon, very few thin particles have been found, and when there is some of them, they are polycrystalline, and smaller planes seem to agglomerate on larger particles. Moreover, SEM analysis of Madagascar graphite also showed small particles, thus it is unsure if the process in argon really does exfoliate the graphite flakes. At the very least, the process in Ar gas does not exfoliate in comparison to the process in air.

As far as the lateral dimensions are concerned, the smallest graphene sheets observed in the case of air as plasmogen gas, smallest particles as the one shown in Fig. 11 were small, mostly around 100 nm, but their thickness is very low.

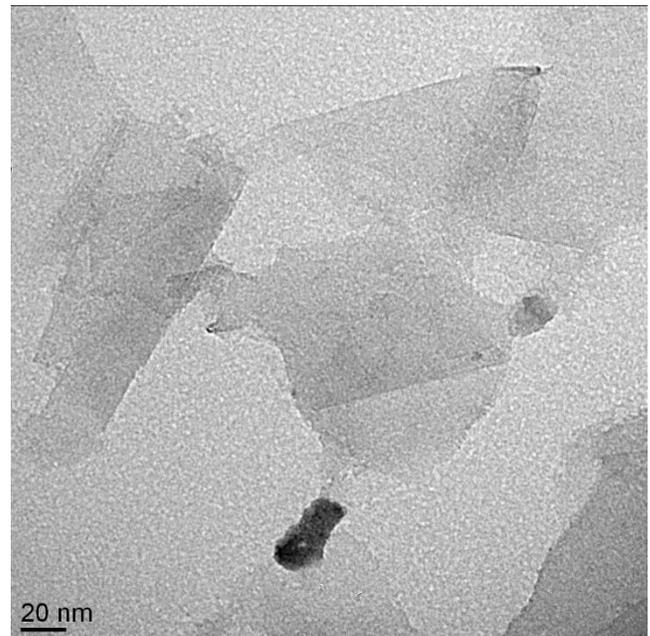

Fig. 11: TEM micrograph of a thin graphene sheet, seen with a contrast diaphragm.

For the thicker particles, as displayed on Fig. 12 for example, their lateral dimensions are between 2 and 20µm. From what is displayed, thicker particles seem large, and thinner particles seem to have small dimensions. Thus, it can be a hypothesis that with this plasma exfoliation process, the thickness and the size of the graphene sheets produced can be correlated.

The effect of low and high current (respectively high and low voltage) can be further analysed with TEM analysis, to know if it also has an effect on the exfoliated graphene sheets. Fig. 12 and Fig. 13 are TEM micrographs coupled with selected area electron diffraction (SAED) to check the crystallinity of the particles.

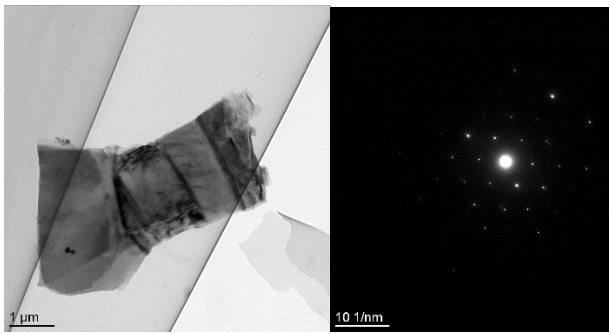

Fig. 12: TEM micrograph (left) of a particle treated with low current and the SAED (right) pattern associated.

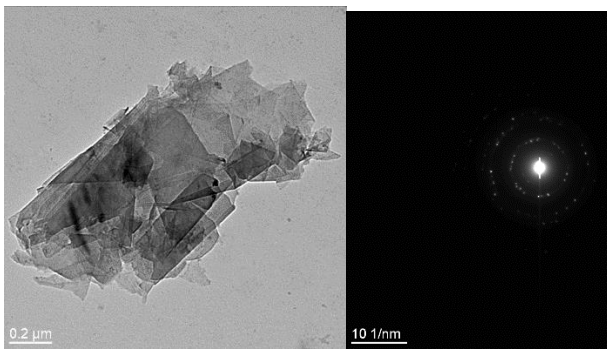

Fig. 13: TEM micrograph (left) of a particle treated with high current and the SAED (right) pattern associated.

From what has been observed with TEM on a small number of particles, the proportion of monocrystalline particles exfoliated with low current seems higher than the proportion exfoliated with high current, but given the small amount of particles observed, no conclusions can be drawn. Furthermore, more defects were found on particles treated with high current, as shown in Fig. 14. Mostly, it consists of cracks and holes, but these have a severe impact on the final properties of the nanomaterial and must be taken into account.

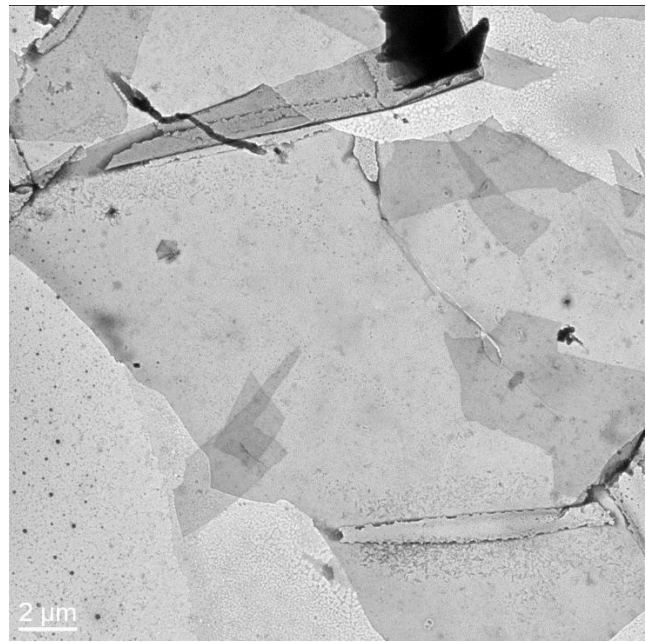

Fig. 14: TEM micrograph of the defects found on a particle treated with high current.

For further research on the particles, holey grids need to be used for TEM analysis, coupled with EELS to confirm or infirm this hypothesis, and obtain a quantitative information regarding the thickness of the graphene sheets.

### III.3.c Functionalization

The functional groups are evaluated after the process, when the graphene sheets have been dried. To avoid combustion of the carbon materials during the TGA, the temperature does not exceed 900 °C in Helium as carrier gas. A mass spectroscopy was coupled with the TGA scale in order to identify the characteristic groups for each mass reduction of the sample.

Recorded TGA curve and its derivative are reported in Fig. 15. A mass reduction is observed at 325 °C. The channels that recorded the higher intensity at 325 °C were channels 18, and 17, showing that water could be trapped inside the material, even after having been dried. These channels are displayed on Fig. 16.

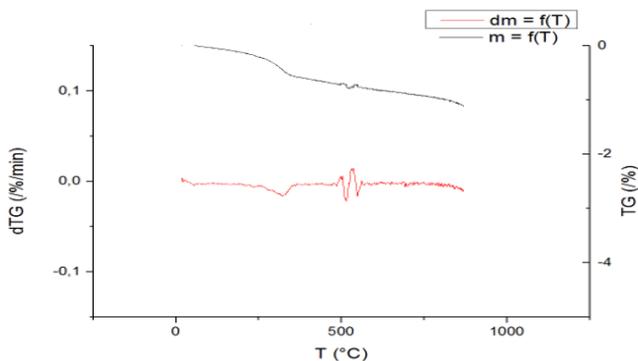

Fig. 15: TGA curve for obtained graphene sheets (black) and its derivative (red).

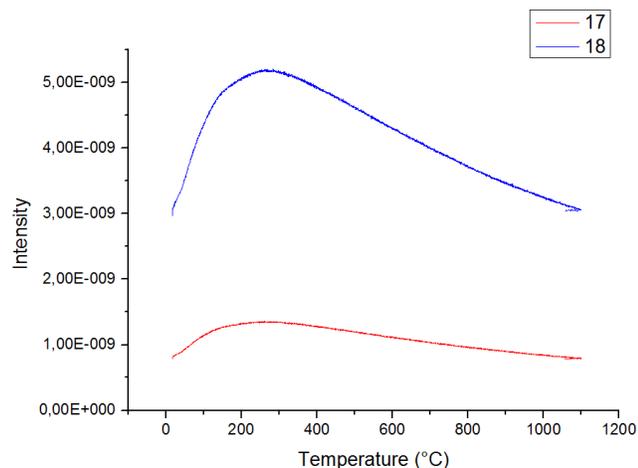

Fig. 16: Channels observed in mass spectrometer during TGA, channels 17 (red) and 18 (blue).

In the same way, Raman spectroscopy can be used to qualitatively know the modifications induced on the material, which mostly consist of degradation. Mainly, the $I_D/I_G$ ratio lets us know if defects have been added to the $sp^2$ structure of graphene. On the graphite, which Raman spectra is recorded on Fig. 17, the ratio can be calculated, and it is $I_D/I_G$ = 0.16. That means that the structure of Madagascar graphite does not show many modifications, given its low ratio. Concerning the modification of graphene, this ratio is the most important to look at.

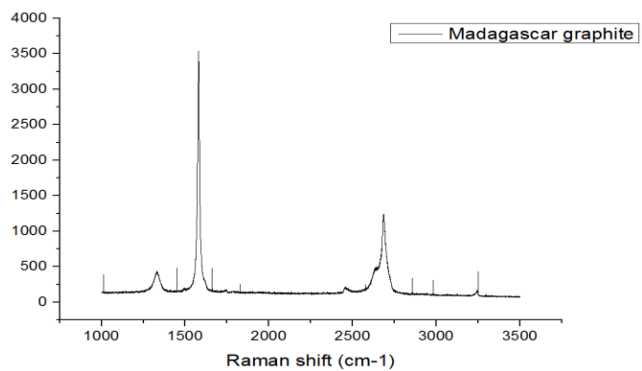

Fig. 17: Raman spectrum of Madagascar graphite.

This ratio is to be compared to the one observed for the exfoliated material. That would inform if defects have been incorporated to the graphene sheets. On Fig. 18 is displayed the Raman spectra of the graphene sheets obtained. Again, the $I_D/I_G$ ratio can be calculated, and we found $I_D/I_G$ = 0.42. However, the mean value of this ratio seems to be near 0.26 in mean. That shows that the process induces defects in the exfoliated sheets. The defects can be of several types, $sp^3$ defects meaning that the sheets have been functionalized, vacancies, or grain boundaries [37].

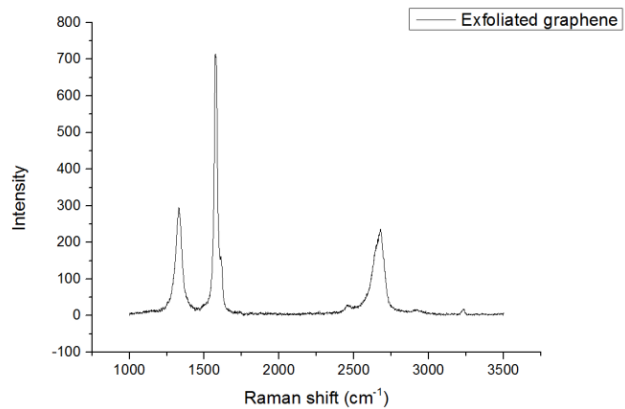

Fig. 18: Raman spectrum of exfoliated graphene sheets.

Furthermore, the intensity ratio $I_D/I_{D'}$ can inform on the nature of the defects. In our case, the ratio $I_D/I_{D'}$= 1.4 for Madagascar graphite, and $I_D/I_{D'}$ = 2 for the exfoliated graphene sheets. This means that more defects, most probably vacancy-like defects due to high current, or more $sp^3$ defects are found on the exfoliated sheets.

III.4 Plasma-activated water diagnosis

Plasma-activated water defines the solution obtained after the process, because the solution is a lot different than its original state.

Fig. 19 shows the decrease of the pH of the solution with process time. As can be seen, in air the final pH value is reached in the beginning of the process, at 10 minutes. However, as seen before for the current kinetics, the pH of the solution does not change during the process in argon.

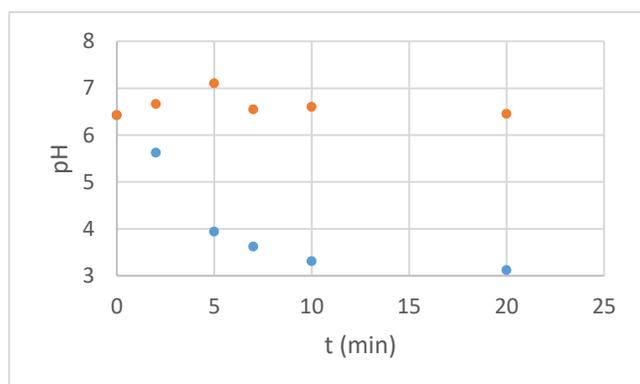

Fig. 19: Evolution of the pH with the process time in air (blue) and argon (orange).

This process was conducted without graphite flakes in the solution, to obtain information about the plasma-activated water only. Fig. 20 shows the electrical conductivity of the plasma-activated water during the process. As expected, it increases as the maximal current during the impulse discharge. In air, the maximal value is reached at 20 minutes, while in argon the electrical conductivity does not change during the process.

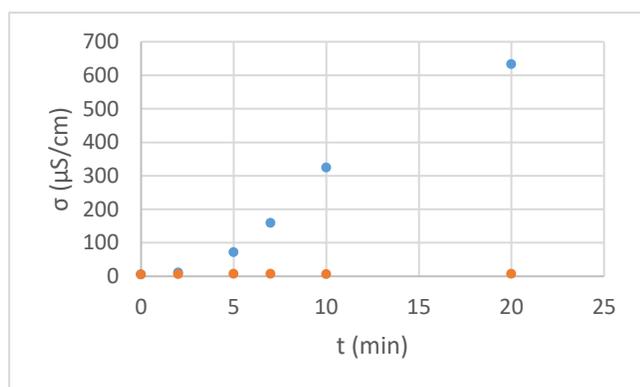

Fig. 20: Evolution on the electrical conductivity of the solution with the process time in air (blue) and argon (orange).

Additionally, a strange behaviour of the plasma-activated water has been observed, it is shown on Fig. 21. When the temperature increases, the electrical conductivity of the solution decreases, while electrolytes containing ions usually see their conductivity increase with temperature.

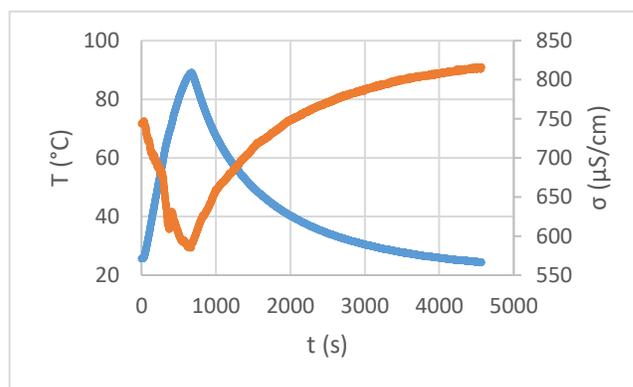

Fig. 21: Evolution of the temperature (blue) and conductivity (orange) when heating plasma-activated water.

Moreover, as can be seen on Fig. 21, a reaction occurs at 70 °C, changing the conductivity of the solution. After heating up the plasma-activated water up to 90 °C and letting it cool down, its conductivity is not the same as before. With more heating cycles, the conductivity at room temperature keeps increasing.

## IV. DISCUSSION

IV.1 Plasma-liquid interface and species in solution

As said in the previous part, a plasma is a very reactive and changing environment, in which many chemical reactions may occur. In this reactive environment are found electrons and ions, with a high kinetic energy and high momentum. This energy can induce reactions in the gas phase and in the plasma-liquid interface.

Plasma-liquid interface is still a new area of research, and it has not been studied nor understood enough. In our process, the most difficult part to understand probably is what happens at this interface. First studies about this interface suggest that the kinetic energy is the determinant factor to have a cascaded liquid-based chemistry [41]. In our case with cathodic polarisation, electrons and anions are going from the cathode to the surface of the liquid with an approximate kinetic energy of $6.10^{-12}$ kg.cm².s⁻².

Still, some species have been identified and sometimes quantified in the plasma-activated water, depending on the gas used as carrier gas. These identified species have been reported in Table 1.

Table 1: Species identified in the plasma-activated water depending on the working gas.

| Species | Air | Argon |
|---|---|---|
| $H_2O_2$ | [21, 22, 23, 24, 25, 26, 27, 28] | [29, 30, 31] |
| °OH | [22, 28] | [29, 30, 32] |
| ONOOH/ONOO$^-$ | [28, 33, 34] | X |
| $O_2$°$^-$ | [22] | [30] |
| $^1O_2$ | [22] | [29] |
| $NO_2^-$ | [21, 25, 26, 28, 35, 36] | X |
| $NO_3^-$ | [21, 24, 25, 26, 28, 35, 36] | X |
| $O_3$ | [27, 28] | X |
| $NO_2$° | [28] | X |
| $HNO_2$ | [27] | X |

As expected, no species containing nitrogen were found when using pure argon. However, *Zhang et al.* [39] results show that these species can be formed during the process when the argon contains 10% N2. Moreover, they found no decrease in pH when using argon as working gas, but when using argon with 10% N2 the decrease in pH is the same as when using air as working gas. Thus, the decrease in pH seems to be related mostly to the presence of nitrogen in the working gas. *Ikawa et al.* [40] related that the presence of N2 conducts to generation of NOx$^-$ and their corresponding acids are responsible for the acidification of the water by plasma treatments. These results are in line with our explanation of the decrease in pH.

Acidic dissolution of the immersed electrode is supposed not to play a role in this process, given the fact that the carbon electrode does not contain transition metal, thus unable to release transition metal ions in solution. Therefore, the pH kinetics is not changed by the presence of the immersed carbon electrode.

IV.2 Link between pH, current, conductivity, temperature

Our process kinetics results imply that there is a link between pH, conductivity, current, and temperature of the solution over time.

First of all, the first event occurring during the process in air gas is the decrease of pH, happening in the first 10 minutes, reaching a plateau of an approximately 3.20 value. After that, simultaneously, we observe an increase in conductivity, current passing through the solution, and an increase of the liquid temperature.

The decrease of pH being mainly caused by the increase of nitrogen-based ions, the concentration of ions in the solution increases. However, the conductivity increases after the plateau of pH is reached. It is supposed that when the plateau is reached, no more NOx$^-$ are produced, but rather other species containing nitrogen, such as reactive nitrogen species as seen in Table 1. Their effect on conductivity could be higher than those of other species in solution.

The Joule heating ($P=R*I^2$) is the source of water heating in the process. The current passing through the liquid is directly impacted by the solution conductivity, thus increasing. However, the resistance decreases as well, since it can be considered as the sum of the resistance of the liquid and the resistance of the gas in which the plasma is generated. The resistance of air is $6,5.10^{18} \Omega$, which is 5 times higher than air saturated with water vapour. As for the liquid, resistance of distilled water is 400 Ω. As the equivalent resistance is the sum of the two values, the resistance of the system is divided by 5 during the process while the current is multiplied by 150. Thus, Joule heating increases with process time, meaning that more thermal energy is released in the liquid, heating the solution

As presented, using argon as working gas does not increase any of the four parameters reported here. No nitrogen-based ions are formed, then no significant change in pH is detected, the current passing through the solution slightly increases but remains very low, around 30 mA, and the heating and increase of conductivity is really low. However, a hydrophilic behaviour of the graphite flakes has been noticed.

First experiments with pure N2 gas as plasmogen reveal that pH decreases, and the current kinetics are similar to those reported in air, only slower. A plateau of 2.8 A has been reached, and a pH of 3.10 has been measured, revealing that the acidity of the solution seems to mostly depend on the presence of nitrogen-based ions, as observed in previous studies.

IV.3 Proposed mechanism for exfoliation

Currently, this exfoliation mechanism is still unknown, even if we have some information about what is needed to exfoliate. Based on what is reported in results part, the plasma is needed to exfoliate, but not any working gas implies exfoliation. Thus, the species formed at the plasma-liquid interface should have a key role.

Moreover, exfoliation has been observed during the first 7 minutes of the process as well as during the last 7 minutes, which points out that the current or pH should not have an influence. The last element to take into account is that graphite flakes not neighbouring the point of plasma impact seem to be unchanged, meaning that they do not disperse in solution nor are fragmented.

Most probably, at the moment when the plasma encounters the liquid surface, a shock wave is generated with a given energy, which could be enough to weaken some Van der Waals bonds at the edge of the graphite flake. Then, a chemical species formed during the air process such as O₂°⁻, which atomic bond distance is smaller than for O2, could intercalate in this weakened place between two layers, thus beginning the exfoliation.

IV.4 Proposed mechanism for functionalization

The functionalization was presented as a possible approach to explain the hydrophilic nature of the graphite flakes after having been hit by the plasma discharge. If this is the case, then the functions added to the graphite are composed of oxygen and hydrogen coming from the dissociation of $H_2O$, and nitrogen coming from the gas phase.

Thus, large graphene sheets in high concentration would mean a very stable suspension. Two possibilities come to mind as to why the flakes become hydrophilic, it can be because they are functionalized, but it also can be because their surface are damaged in some way, thus making it less planar, and decrease its hydrophobic nature.

As shown in the results part concerning the dispersion of the particles in solution, the particles have to be near the impact of the plasma the become hydrophilic, and this behaviour seems to be due to the functionalization of the graphite flakes.

This functionalization happens on particles that are near the impact of the plasma with the surface of the solution. The same observation is made independently of the gas used (air, argon, nitrogen), letting think that the functional groups are made of the particles that are already found in the solution. Since distilled water is used, functional groups should be derived from $H_2O$.

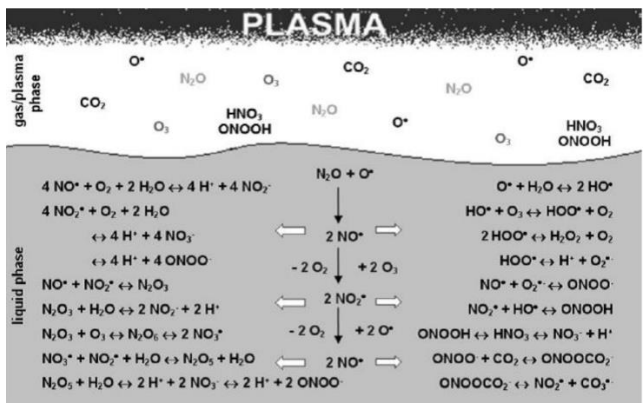

Fig. 22: Assumption of chemical reactions in plasma-treated liquids[38].

As seen on Fig. 22, a lot of reactions occur in the interface between the plasma and the liquid, forming a lot of different species as reported previously in Table 1. However, since the functionalization in our case was observed in argon too, we assume that the breaking of the $H_2O$ molecule forms free radicals such as °OH. Then, since these radicals are highly unstable, they tend to recombine with what they find around them. Given the number of dissociations that occur in the interface at the same time, it is reasonable to say that some radicals are reacting with the graphite flakes near them, functionalizing them, and changing their behaviour to make them hydrophilic.

IV.5 Hypotheses for the conductivity of the water

As presented in the results part, a remarkable property of the plasma-treated water has been noticed. From what is known, the plasma-treated water should act as an electrolyte, since it is a solution that contains ions in a relatively high concentration. Such a solution presents a conductivity that increases with temperature, because the ions mobility is increasing meaning they conduct current more easily, contrarily to our solution. Moreover, a solution in which no reaction occurs has a given conductivity at a temperature, no matter the heat treatment.

Thus, it is clear that a reaction occurs in our solution, at approximately 70° C, irreversibly increasing its conductivity. Moreover, an unknown species reversibly decreases the solution's conductivity to counteract its increase due to the ions.

That means that two different unknown reactions occur when heating the plasma-activated water. The reversible reaction could be an equilibrium reaction between an ion and its corresponding acid, decreasing the concentration of ions in solution thus decreasing the conductivity, and another hypothesis is that when heating, ions have a higher mobility and combine with opposing charged ions, which will not contribute to the conductivity.

For the irreversible reaction occurring at 70° C, the hypothesis of the decomposition of neutral species such as $H_2O_2$ can explain this change in conductivity. $H_2O_2$ not being very stable, and the only neutral species known to be formed during this process, its decomposition is the most probable explanation for this reaction.

IV.6 Possibilities for gas mixtures to optimize the process

For now, only three working gases have been used: air, argon and nitrogen. Using argon, we saw that the working gas has a vast impact on the process, mainly for the exfoliation part. While no to little exfoliation was observed in the case of argon as working gas, a huge difference was found in the thickness of the particles after air treatment.

This guides us to think that gas mixtures could be a good alternative to these gases, as it could easily change the process kinetics, and allow to understand better the process mechanisms.

**CONCLUSIONS**

In summary, we have synthesized graphene sheets from graphite flakes using plasma discharge between an electrode and the surface of distilled water. This process activates the water, exfoliates and functionalizes the graphite flakes. With a carbon electrode immersed in the water, no contamination or other carbonated species are found in solution. The graphene sheets produced were hundreds of nanometers in size, however a precise measure of the thickness of the graphene sheets with electron energy loss spectroscopy (EELS) was not conducted. The produced graphene sheets are monocrystalline, and functionalized to be used in polymer matrixes. Further studies need to be done to evaluate the optimal electrical parameters, the influence of the solvent, and to analyse the activated water.

ACKNOWLEDGMENTS
Research was supported by the Institut Jean Lamour.

REFERENCES

[1] Lee, C., Wei, X., Kysar, J. and Hone, J. (2008). Measurement of the Elastic Properties and Intrinsic Strength of Monolayer Graphene. *Science*, 321(5887), pp.385-388.

[2] Stoller, M., Park, S., Zhu, Y., An, J. and Ruoff, R. (2008). Graphene-Based Ultracapacitors. *Nano Letters*, 8(10), pp.3498-3502.

[3] Nair, R., Blake, P., Grigorenko, A., Novoselov, K., Booth, T., Stauber, T., Peres, N. and Geim, A. (2008). Fine Structure Constant Defines Visual Transparency of Graphene. *Science*, 320(5881), pp.1308-1308.

[4] Reina, A., Jia, X., Ho, J., Nezich, D., Son, H., Bulovic, V., Dresselhaus, M. and Kong, J. (2009). Large Area, Few-Layer Graphene Films on Arbitrary Substrates by Chemical Vapor Deposition. *Nano Letters*, 9(1), pp.30-35.

[5] Balandin, A., Ghosh, S., Bao, W., Calizo, I., Teweldebrhan, D., Miao, F. and Lau, C. (2008). Superior Thermal Conductivity of Single-Layer Graphene. *Nano Letters*, 8(3), pp.902-907.

[6] Van Noorden, R. (2011). Chemistry: The trials of new carbon. *Nature*, 469(7328), pp.14-16.

[7] Chandra, V., Park, J., Chun, Y., Lee, J., Hwang, I. and Kim, K. (2010). Water-Dispersible Magnetite-Reduced Graphene Oxide Composites for Arsenic Removal. *ACS Nano*, 4(7), pp.3979-3986.

[8] Lee, W., Park, J., Kim, Y., Kim, K., Hong, B. and Cho, K. (2011). Control of Graphene Field-Effect Transistors by Interfacial Hydrophobic Self-Assembled Monolayers. *Advanced Materials*, 23(30), pp.3460-3464.

[9] Novoselov, K., Geim, A., Morozov, S., Jiang, D., Katsnelson, M., Grigorieva, I., Dubonos, S. and Firsov, A. (2005). Two-dimensional gas of massless Dirac fermions in graphene. *Nature*, 438(7065), pp.197-200.

[10] Kim, K., Zhao, Y., Jang, H., Lee, S., Kim, J., Kim, K., Ahn, J., Kim, P., Choi, J. and Hong, B. (2009). Large-scale pattern growth of graphene films for stretchable transparent electrodes. *Nature*, 457(7230), pp.706-710.

[11] Li, X., Cai, W., An, J., Kim, S., Nah, J., Yang, D., Piner, R., Velamakanni, A., Jung, I., Tutuc, E., Banerjee, S., Colombo, L. and Ruoff, R. (2009). Large-Area Synthesis of High-Quality and Uniform Graphene Films on Copper Foils. *Science*, 324(5932), pp.1312-1314.

[12] Sun, Z., Yan, Z., Yao, J., Beitler, E., Zhu, Y. and Tour, J. (2010). Growth of graphene from solid carbon sources. *Nature*, 468(7323), pp.549-552.

[13] Sutter, P., Flege, J. and Sutter, E. (2008). Epitaxial graphene on ruthenium. *Nature Materials*, 7(5), pp.406-411.

[14] Grande, L., Chundi, V., Wei, D., Bower, C., Andrew, P. and Ryhänen, T. (2012). Graphene for energy harvesting/storage devices and printed electronics. *Particuology*, 10(1), pp.1-8.

[15] Karuwan, C., Sriprachuabwong, C., Wisitsoraat, A., Phokharatkul, D., Sritongkham, P. and Tuantranont, A. (2012). Inkjet-printed graphene-poly(3,4-ethylenedioxythiophene):poly(styrene-sulfonate) modified on screen printed carbon electrode for electrochemical sensing of salbutamol. *Sensors and Actuators B: Chemical*, 161(1), pp.549-555.

[16] Cheng, M., Yang, R., Zhang, L., Shi, Z., Yang, W., Wang, D., Xie, G., Shi, D. and Zhang, G. (2012). Restoration of graphene from graphene oxide by defect repair. *Carbon*, 50(7), pp.2581-2587.

[17] Turner, P., Hodnett, M., Dorey, R. and Carey, J. (2019). Controlled Sonication as a Route to in-situ Graphene Flake Size Control. *Scientific Reports*, 9(1).

[18] Ostrikov, K., Neyts, E. and Meyyappan, M. (2013). Plasma nanoscience: from nano-solids in plasmas to nano-plasmas in solids. *Advances in Physics*, 62(2), pp.113-224.

[19] Volotskova, O., Levchenko, I., Shashurin, A., Raitses, Y., Ostrikov, K. and Keidar, M. (2010). Single-step synthesis and magnetic separation of graphene and carbon nanotubes in arc discharge plasmas. *Nanoscale*, 2(10), p.2281.

[20] Sano, N., Wang, H., Alexandrou, I., Chhowalla, M., Teo, K., Amaratunga, G. and Iimura, K. (2002). Properties of carbon onions produced by an arc discharge in water. *Journal of Applied Physics*, 92(5), pp.2783-2788.

[21] Liu, F., Sun, P., Bai, N., Tian, Y., Zhou, H., Wei, S., Zhou, Y., Zhang, J., Zhu, W., Becker, K. and Fang, J. (2010). Inactivation of Bacteria in an Aqueous Environment by a Direct-Current, Cold-Atmospheric-Pressure Air Plasma Microjet. *Plasma Processes and Polymers*, 7(3-4), pp.231-236.

[22] Sun, P., Wu, H., Bai, N., Zhou, H., Wang, R., Feng, H., Zhu, W., Zhang, J. and Fang, J. (2011). Inactivation of Bacillus subtilis Spores in Water by a Direct-Current, Cold Atmospheric-Pressure Air Plasma Microjet. *Plasma Processes and Polymers*, 9(2), pp.157-164.

[23] Shainsky, N., Dobrynin, D., Ercan, U., Joshi, S., Ji, H., Brooks, A., Fridman, G., Cho, Y., Fridman, A. and Friedman, G. (2012). Retraction: Plasma Acid: Water Treated by Dielectric Barrier Discharge. *Plasma Processes and Polymers*, 9(6).

[24] Burlica, R., Grim, R., Shih, K., Balkwill, D. and Locke, B. (2010). Bacteria Inactivation Using Low Power Pulsed Gliding Arc Discharges with Water Spray. *Plasma Processes and Polymers*, 7(8), pp.640-649.

[25] Oehmigen, K., Hoder, T., Wilke, C., Brandenburg, R., Hahnel, M., Weltmann, K. and von Woedtke, T. (2011). Volume Effects of Atmospheric-Pressure Plasma in Liquids. *IEEE Transactions on Plasma Science*, 39(11), pp.2646-2647.

[26] Oehmigen, K., Hähnel, M., Brandenburg, R., Wilke, C., Weltmann, K. and von Woedtke, T. (2010). The Role of Acidification for Antimicrobial Activity of Atmospheric Pressure Plasma in Liquids. *Plasma Processes and Polymers*, 7(3-4), pp.250-257.

[27] Shimizu, T., Iwafuchi, Y., Morfill, G. and Sato, T. (2011). Transport Mechanism of Chemical Species in a Pin-water Atmospheric Discharge driven by Negative Voltage. *Journal of Photopolymer Science and Technology*, 24(4), pp.421-427.

[28] Locke, B. and Lukeš, P. (2018). Special issue: Plasma and Liquids. *Plasma Processes and Polymers*, 15(6), p.243-308.

[29] Zhang, Q., Sun, P., Feng, H., Wang, R., Liang, Y., Zhu, W., Becker, K., Zhang, J. and Fang, J. (2012). Assessment of the roles of various inactivation agents in an argon-based direct current atmospheric pressure cold plasma jet. *Journal of Applied Physics*, 111(12), p.1-6.

[30] Jablonowski, H., Bussiahn, R., Hammer, M., Weltmann, K., von Woedtke, T. and Reuter, S. (2015). Impact of plasma jet vacuum ultraviolet radiation on reactive oxygen species generation in bio-relevant liquids. *Physics of Plasmas*, 22(12).

[31] Kurita, H., Nakajima, T., Yasuda, H., Takashima, K., Mizuno, A., Wilson, J. and Cunningham, S. (2011). Single-molecule measurement of strand breaks on large DNA induced by atmospheric pressure plasma jet. *Applied Physics Letters*, 99(19), p.191504.

[32] Okazaki, Y., Wang, Y., Tanaka, H., Mizuno, M., Nakamura, K., Kajiyama, H., Kano, H., Uchida, K., Kikkawa, F., Hori, M. and Toyokuni, S. (2014). Direct exposure of non-equilibrium atmospheric pressure plasma confers simultaneous oxidative and ultraviolet modifications in biomolecules. *Journal of Clinical Biochemistry and Nutrition*, 55(3), pp.207-215.

[33] Oehmigen, K., Winter, J., Hähnel, M., Wilke, C., Brandenburg, R., Weltmann, K. and von Woedtke, T. (2011). Plasma Process. Polym. 10/2011. *Plasma Processes and Polymers*, 8(10), pp.904-913.

[34] Machala, Z., Tarabova, B., Hensel, K., Spetlikova, E., Sikurova, L. and Lukes, P. (2013). Formation of ROS and RNS in Water Electro-Sprayed through Transient Spark Discharge in Air and their Bactericidal Effects. *Plasma Processes and Polymers*, 10(7), pp.649-659.

[35] Pavlovich, M., Ono, T., Galleher, C., Curtis, B., Clark, D., Machala, Z. and Graves, D. (2014). Air spark-like plasma source for antimicrobial NOx generation. *Journal of Physics D: Applied Physics*, 47(50).

[36] Pavlovich, M., Clark, D. and Graves, D. (2014). Quantification of air plasma chemistry for surface disinfection. *Plasma Sources Science and Technology*, 23(6), p.065036.

[37] Eckmann, A., Felten, A., Mishchenko, A., Britnell, L., Krupke, R., Novoselov, K. and Casiraghi, C. (2012). Probing the Nature of Defects in Graphene by Raman Spectroscopy. *Nano Letters*, 12(8), pp.3925-3930.

[38] Jablonowski, H. and von Woedtke, T. (2015). Research on plasma medicine-relevant plasma–liquid interaction: What happened in the past five years?. *Clinical Plasma Medicine*, 3(2).

[39] Zhang, Q., Sun, P., Feng, H., Wang, R., Liang, Y., Zhu, W., Becker, K., Zhang, J. and Fang, J. (2012). Assessment of the roles of various inactivation agents in an argon-based direct current atmospheric pressure cold plasma jet. *Journal of Applied Physics*, 111(12), pp.1-6.

[40] Ikawa, S., Kitano, K. and Hamaguchi, S. (2010). Effects of pH on Bacterial Inactivation in Aqueous Solutions due to Low-Temperature Atmospheric Pressure Plasma Application. *Plasma Processes and Polymers*, 7(1), pp.33-42.

[41] Mariotti, D., Patel, J., Švrček, V. and Maguire, P. (2012). Plasma-Liquid Interactions at Atmospheric Pressure for Nanomaterials Synthesis and Surface Engineering. *Plasma Processes and Polymers*, 9(11-12), pp.1074-1085.